\documentclass[aps,preprintnumbers,epsfig,prl,twocolumn,showpacs,superscriptaddress]{revtex4}
\usepackage{amssymb}
\usepackage{amsmath}
\usepackage{amsopn}
\usepackage{graphics}
\usepackage{graphicx}
\usepackage{epsfig}

\begin{document}

\title{Early out-of-equilibrium beam-plasma evolution}
\author{M.-C. Firpo}
\affiliation{Laboratoire de Physique et Technologie des Plasmas
(CNRS UMR 7648), Ecole Polytechnique, 91128 Palaiseau cedex, France}
\author{A. F. Lifschitz}
\affiliation{Laboratoire d'Optique Appliqu\'{e}e, ENSTA/Ecole
Polytechnique (CNRS UMR 7639), 91761 Palaiseau cedex, France}
\author{E. Lefebvre}
\affiliation{D\'{e}partement de Physique Th\'{e}orique et
Appliqu\'{e}e, CEA/DAM Ile-de-France, BP 12, 91680
Bruy\`{e}res-le-Ch\^{a}tel, France}
\author{C. Deutsch}
\affiliation{Laboratoire de Physique des Gaz et des Plasmas (CNRS
UMR 8578), Universit\'{e} Paris XI, B\^{a}timent 210, 91405 Orsay
cedex, France}
\date{\today}
\preprint{}

\begin{abstract}
We solve analytically the out-of-equilibrium initial stage that
follows the injection of a radially finite electron beam into a
plasma at rest and test it against particle-in-cell simulations. For
initial large beam edge gradients and not too large beam radius,
compared to the electron skin depth, the electron beam is shown to
evolve into a ring structure. For low enough transverse
temperatures, the filamentation instability eventually proceeds and
saturates when transverse isotropy is reached. The analysis accounts
for the variety of very recent experimental beam transverse
observations.
\end{abstract}

 \maketitle

Beam-plasma interactions have recently received some considerable renewed
interest especially in the relatively unexplored regimes of high beam
intensities and high plasma densities. One particular motivation lies in the
fast ignition schemes (FIS) for inertial confinement fusion \cite{Tabak94}.
These should involve in their final stage the interaction of an ignition
beam composed of MeV electrons laser generated at the critical density
surface with a dense plasma target. The exploration of the electron beam
transport into the overdense plasma is essential to assess the efficiency of
the beam energy deposit. In this matter, transverse beam-plasma
instabilities could be particularly deleterious in preventing conditions for
burn to be met. Experimental observations recently undertaken in conditions
relevant to the FIS have either shown some transverse microscopic
filamentation of electron beams \cite{Tatarakis03} or some transverse,
predominantly macroscopic, beam evolution into a ring structure \cite%
{Koch2002,Jung2005} or a superposition of those effects \cite%
{SteinReport2003,Jung2005,Norreys}, with filaments standing out from
a ring structure, in a scenario similar to Taguchi \textit{et al.}'s
numerical simulations \cite{Taguchi2001}. Weibel instability
\cite{Weibel} is commonly invoked to account for these phenomena,
but it is sometimes difficult to find any clear univocal evidence
supporting this. Moreover the fact is that, whereas most theoretical
and some computational studies are devoted to the linear regime of
instabilities originating from current and charge neutralized
equilibria, the physics of the fast ignition is intrinsically
out-of-equilibrium.

In this Letter, we shall consider the out-of-equilibrium initial value
dynamical problem taking place when a radially inhomogeneous electron
forward current is launched into a plasma and is still not current
compensated. We shall focus on this early stage where collisions may be
neglected. Ions will be assumed to form a fixed neutralizing background. In
order to simplify both the analysis and the numerical PIC computations, we
shall consider the system to be infinite along the beam direction $z$. We
shall remove any $z$ dependance by assuming also that plasma density $n_{pe}$
is uniform and constant. At time $t=0$, a relativistic electron beam is
switched on in the plasma.

Maxwell equations are linear and can thus be solved for all time to
give the electromagnetic fields as functions of the sources, namely
beam and plasma current densities, $\mathbf{j}_{b}$ and
$\mathbf{j}_{pe}$. We get $\mathop{\rm rot}\nolimits\mathbf{B}=\mu _{0}(\mathbf{j}_{pe}+\mathbf{j}%
_{b})+1/c^{2}\partial \mathbf{E}/\partial t$ and $\mathop{\rm
rot}\nolimits\mathbf{E}=-\partial \mathbf{B}/\partial t$. The
electron plasma current $\mathbf{j}_{pe}=-en_{pe}\mathbf{v}_{pe}$ is
initially vanishing and may be approximated by linear fluid theory
in the initial stage yielding
\begin{equation}
\frac{\partial \mathbf{j}_{pe}}{\partial t}=\varepsilon _{0}\omega _{pe}^{2}%
\mathbf{E},  \label{linear_jp}
\end{equation}%
with $\omega _{pe}=\sqrt{n_{pe}e^{2}/m_{e}\varepsilon _{0}}$ the plasma
pulsation. We Fourier decompose any field $g$ through $g(r,\theta
,t)=\sum_{m}g^{(m)}(r,t)\exp \left( im\theta \right) $ and proceed to a
Laplace transform in time $\hat{g}^{(m)}\left( r,s\right) =\int_{0}^{\infty
}e^{-st}g^{(m)}\left( r,t\right) dt$. Eliminating the electric field
components, Maxwell equations in cylindrical geometry yield inhomogeneous
wave equations with sources for the magnetic field components. Introducing
the operator $\mathcal{L}_{n}[\mu ]$ such that
\begin{equation}
\mathcal{L}_{n}[\mu ]y\equiv \frac{1}{r}\frac{d}{dr}\left( r\frac{dy}{dr}%
\right) +\left( \mu ^{2}-\frac{n^{2}}{r^{2}}\right) y,  \label{def_Ln}
\end{equation}%
defining $\sigma \equiv \sqrt{s^{2}+\omega _{pe}^{2}}/c$ and neglecting the
initial values of the e.m. fields, the wave equations read
\begin{eqnarray}
\text{for }m &=&0\text{, }\mathcal{L}_{1}\left[ i\sigma \right] \hat{B}%
_{\theta }^{(0)}=\mu _{0}\frac{\partial \hat{\jmath}_{bz}^{(0)}}{\partial r},
\label{Btheta_0} \\
\text{for }m &\neq &0\text{, }i\mathcal{L}_{m}\left[ i\sigma \right] \left( r%
\hat{B}_{r}^{(m)}\right) =m\mu _{0}\hat{\jmath}_{bz}^{(m)},
\label{rBr_mnon0}
\end{eqnarray}%
with, for any $m$,
\begin{equation}
\mathcal{L}_{m}\left[ i\sigma \right] \hat{B}_{z}^{(m)}=\frac{\mu _{0}}{r}%
\left[ im\hat{\jmath}_{br}^{(m)}-\frac{\partial }{\partial r}\left( r\hat{%
\jmath}_{b\theta }^{(m)}\right) \right] .  \label{Bz_annym}
\end{equation}%
Let us make the following general statements: Because they are linear,
Maxwell equations do not enable spectral changes. If the beam is
sufficiently weak, so that the fluid approximation for the bulk plasma
remains approximately valid, mode transfers will originate from the beam
particles equations of motion. Consequently, if the initial beam is
rigorously both rotationally and axially invariant (on $m=0$), it will
remain so for all times. Then there are no sources to feed the triad ($B_{r}$%
, $B_{z}$, $E_{\theta }$) that remains vanishingly small. This is an
invitation to focus first on the $m=0$ evolution.

We consider initial beam density and velocity of $n_{b0}(r)=n_{b0}%
\bar{n}(\tilde{r})$ and $v_{b0z}(r)=v_{b0z}\bar{v}(\tilde{r})$, with $\tilde{%
r}=r/r_{b}$. Let us introduce here the beam radius $r_{b}$
\cite{note1}, the electron skin depth $\lambda _{s}\equiv c/\omega
_{pe}$, their ratio $\eta \equiv r_{b}/\lambda _{s}$, and let us
define $\alpha =n_{b0}/n_{pe}$, $\beta _{0}=v_{b0z}/c$ and the
initial relativistic Lorentz factor $\gamma _{0}=\left(
1-v_{b0z}^{2}/c^{2}\right) ^{-1/2}=\left( 1-\beta
_{0}^{2}\bar{v}^{2}\right) ^{-1/2}=\gamma _{0}(\tilde{r})$. The
Green function $g(r\mid a)$ \cite{Duffy} solving
$\mathcal{L}_{1}\left[ i\sigma \right] g=-\delta \left( r-a\right) $
is readily computed as $g(r\mid a)=I_{1}\left( \sigma r^{<}\right)
K_{1}\left( \sigma r^{>}\right) $ with $r^{<}=\min (r,a)$ and
$r^{>}=\max
(r,a)$. The general solution of Eq. (\ref{Btheta_0}) is then%
\begin{equation}
\hat{B}_{\theta }^{(0)}(r,s)=-\mu _{0}\int_{0}^{+\infty }aI_{1}(\sigma
r^{<})K_{1}(\sigma r^{>})\frac{\partial \hat{\jmath}_{bz}^{(0)}}{\partial a}%
da.  \label{solu_Btheta0}
\end{equation}%
Let us consider the response $\hat{B}_{\theta 1}^{(0)}$ to the initial beam
current $j_{b0z}^{(0)}=-en_{b0}v_{b0z}\bar{j}H(t)$ that is switched on at
time $0$. Here $H$ denotes the Heaviside step function and $\bar{j}(\tilde{r}%
)=\bar{n}(\tilde{r})\bar{v}(\tilde{r})$. This gives $\hat{\jmath}%
_{0bz}^{(0)}(\tilde{r},s)=-en_{b0}v_{b0z}\bar{j}(\tilde{r})/s$. Eq. (\ref%
{Btheta_0}) admits then a solution in separate variables. This makes Laplace
inversion easier giving
\begin{equation}
\frac{e}{m_{e}}B_{\theta 1}^{(0)}\left( \tilde{r},t\right) =\alpha \beta
_{0}\omega _{pe}\eta F(\tilde{r},\eta )\left[ 1-\cos \left( \omega
_{pe}t\right) \right]  \label{solu_Btheta0_order1_time}
\end{equation}%
where the radial information is contained into
\begin{equation}
F(\tilde{r},\eta )=K_{1}(\eta \tilde{r})\int_{0}^{\tilde{r}}uI_{1}(\eta u)%
\bar{j}^{\prime }(u)du+I_{1}(\eta \tilde{r})\int_{\tilde{r}}^{\infty
}uK_{1}(\eta u)\bar{j}^{\prime }(u)du.  \label{defi_F_radial}
\end{equation}%
Then, integrating $\partial E_{z}^{(0)}/\partial r=\partial B_{\theta
}^{(0)}/\partial t$ and using Eq. (\ref{solu_Btheta0_order1_time})
immediately gives
\begin{equation}
\frac{e}{m_{e}}E_{z1}^{(0)}\left( \tilde{r},t\right) =\alpha \beta _{0}\eta
^{2}c\omega _{pe}\sin \left( \omega _{pe}t\right) \int_{\infty }^{\tilde{r}%
}F(u,\eta )du.  \label{solu_Ez_order1}
\end{equation}%
Finally, the $m=0$ radial electric field component satisfies the wave
equation $\partial _{tt}E_{r}^{(0)}+\omega
_{pe}^{2}E_{r}^{(0)}=-1/\varepsilon _{0}\partial _{t}j_{br}^{(0)}$. It is
easy to check that its initial behavior is given by
\begin{equation}
\frac{e}{m_{e}}E_{r}^{(0)}(r,t)=\alpha \omega _{pe}^{2}\bar{n}(\tilde{r}%
)\int_{0}^{t}\cos \left[ \omega _{pe}(t-\tau )\right] v_{b1r}^{(0)}(r,\tau
)d\tau .  \label{eqEr}
\end{equation}%
For the problem under consideration, the beam to plasma density ratio $%
\alpha $ is typically a small parameter. Eq. (\ref{eqEr}) will be second
order in $\alpha $.

\begin{figure}[tbph]
\begin{center}
\resizebox{8.0cm}{4.5cm}{\includegraphics[bb=102 0 403
208]{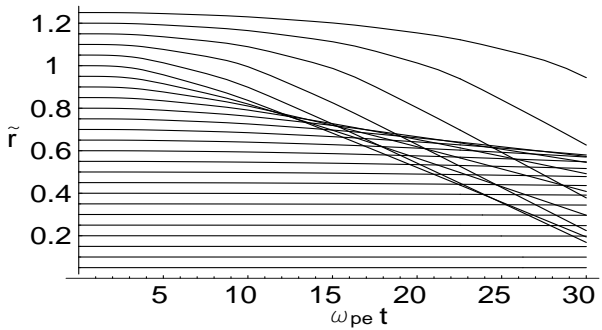}}
\end{center}
\caption{Radial trajectories of test electrons for an initial beam
profile given by Eq. (\ref{flat-profile}) with $\eta=\kappa=10$.}
\label{plot-PartTest}
\end{figure}

We now wish to compute the beam evolution under the previous self-fields (%
\ref{solu_Btheta0_order1_time}), which was ignored in previous
studies \cite{Kuppers73}. Let us assume that the beam can be treated
as a cold fluid and write the $m=0$ fluid equations, dropping the
$m=0$ superscripts,
\begin{eqnarray}
\frac{\partial n_{b}}{\partial t}+\frac{1}{r}\frac{\partial }{\partial r}%
\left( rn_{b}v_{br}\right) &=&0,  \label{eq_cons-matiere} \\
\left( \frac{\partial }{\partial t}+v_{br}\frac{\partial }{\partial r}%
\right) \left( \gamma v_{br}\right) &=&-\frac{e}{m_{e}}E_{r}+\frac{e}{m_{e}}%
v_{bz}B_{\theta },  \label{eq_mouv_radia} \\
\left( \frac{\partial }{\partial t}+v_{br}\frac{\partial }{\partial r}%
\right) \left( \gamma v_{bz}\right) &=&-\frac{e}{m_{e}}E_{z}+\frac{e}{m_{e}}%
v_{br}B_{\theta }.  \label{eq_mouv_axial}
\end{eqnarray}%
Let us write $n_{b}=n_{b0}+n_{b1}(r,t)$, $v_{br}=v_{b1r}(r,t)$ and $%
v_{bz}=v_{b0z}+v_{b1z}(r,t)$. Considering $\alpha $ as a small parameter, we
get a natural hierarchy: first order terms should be of order $\alpha $,
second order terms of order $\alpha ^{2}$ and so on. Let us explicit first
order fluid equations. The radial electric contribution being negligible (%
\ref{eqEr}), Eq. (\ref{eq_mouv_radia}) gives%
\begin{equation}
\gamma _{0}\frac{\partial v_{b1r}}{\partial t}=\alpha \beta _{0}^{2}\eta
c\omega _{pe}\bar{v}(\tilde{r})F(\tilde{r},\eta )\left[ 1-\cos \left( \omega
_{pe}t\right) \right]  \label{eqmouv_1order_r}
\end{equation}%
which yields, using $v_{b1r}(r,t=0)=0$,%
\begin{equation}
\beta_{1r}=\frac{v_{b1r}(\tilde{r},t)}{c}=\alpha \eta \beta _{0}^{2}\frac{\bar{v}(%
\tilde{r})F(\tilde{r},\eta )}{\gamma _{0}(\tilde{r})}\left[ \omega
_{pe}t-\sin \left( \omega _{pe}t\right) \right] .  \label{solu_vb1r}
\end{equation}%
The first order conservation equation is%
\begin{equation}
\frac{\partial n_{b1}}{\partial t}+\frac{1}{r}\frac{\partial }{\partial r}%
\left( rn_{b0}(r)v_{b1r}\right) =0.  \label{eq_cons_mat_order1}
\end{equation}%
Using (\ref{solu_vb1r}), this gives, with $n_{b1}(\tilde{r},t=0)=0$,
\begin{equation}
\frac{n_{b1}(\tilde{r},t)}{n_{b0}}=\alpha \beta _{0}^{2}\bar{n}_{1}(\tilde{r}%
)\left( \frac{1}{2}\omega _{pe}^{2}t^{2}+\cos \left( \omega _{pe}t\right)
-1\right) ,  \label{solu_nb1}
\end{equation}%
where $\bar{n}_{1}$ is the radial function
\begin{equation}
\bar{n}_{1}(\tilde{r})=-\frac{1}{\tilde{r}}\frac{\partial }{\partial \tilde{r%
}}\left[ \frac{\tilde{r}\bar{j}(\tilde{r})F(\tilde{r},\eta )}{\gamma _{0}(%
\tilde{r})}\right] .  \label{definition_n1}
\end{equation}%
\begin{figure}[htbp]
\begin{center}
\resizebox{8.0cm}{8.0cm}{\includegraphics[bb=258 12 708
521]{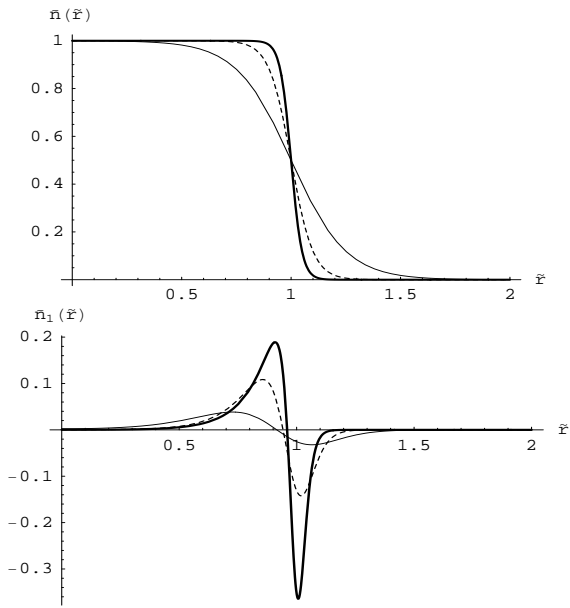}}
\end{center}
\caption{Functions $\bar{n}(\tilde{r})=\left( \tanh \left[ 2%
\protect\kappa (1-\tilde{r})\right] +1\right) /2$ and
$\bar{n}_{1}(\tilde{r}) $ for $\protect\eta =10$ and $\protect\kappa
=10$ (bold), $\protect\kappa =5$ (dashed line) and $\protect\kappa
=2$ (plain line) for a monokinetic beam. } \label{plot-nb0}
\end{figure}
Eq. (\ref{solu_nb1}) shows that $n_{b1}$ has a secular behavior.
Thus the present analysis breaks when
$n_{b1}/n_{b0}=\mathcal{O}(1)$, namely roughly for $\omega
_{pe}t\sim \alpha ^{-1/2}$. In order to put this more precisely, we
shall study the radial behavior $\bar{n}_{1}$. Let us consider some
initially monokinetic beam ($\bar{v}=1$) having density functions of
the form
\begin{equation}\bar{n}(\tilde{r})= \left( \tanh \left[
2\kappa (1-\tilde{r})\right] +1\right)/2. \label{flat-profile}
\end{equation}
This enables the study of the influence of the beam edge gradients
as $\kappa =-\bar{n}^{\prime }(\tilde{r}=1)$. Fig. \ref{plot-nb0}
displays various $\bar{n}$ profiles and their associated first order
perturbations $\bar{n}_{1}$. This shows that, within a given initial
lapse of time, the natural evolution of the system tends to increase
the beam density around some radius below $r_{b}$. This favors the
formation of a ring structure for the beam density, that is all the
sharper and all the closer to $r_{b}$ that the initial radial beam
gradients are high. The emergence of this beam ring formation can be
already inferred from the time evolution of test electrons within
the azimuthal magnetic self-field corresponding to some given
initial profile $\bar{n}$ of the form (\ref{flat-profile}) as shown
in Fig. \ref{plot-PartTest}. The caustics pattern signals there a
cusp formation in the radial beam density. It is important to note
that, due to Eq. (\ref{solu_Btheta0}), this behavior is very
dependent on the initial beam profile. Indeed, we observed no such
caustics pattern nor any visible evolution towards a ring structure
for smooth Gaussian initial beam profile, but rather radial focusing
prior to the filamentation onset.

\begin{figure}[tbph]
\begin{center}
\rotatebox{-90}{\includegraphics[width=6cm]{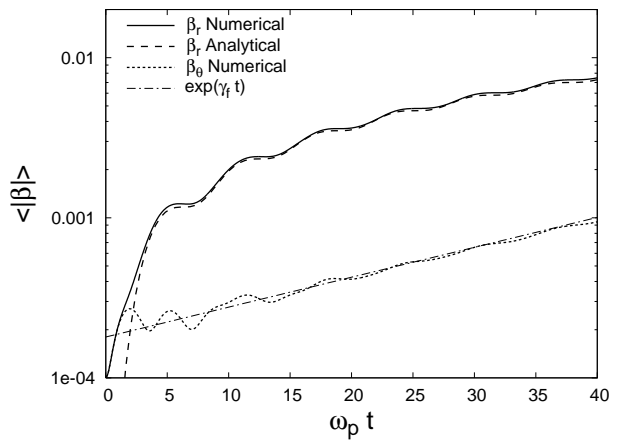}}
\end{center}
\caption{Early time evolution of the modulus of the radial average
of the transversal beam velocity ($\beta_{r}=v_{r}/c$ and
$\beta_{\theta}=v_{\theta}/c$) in lin-log scale for $\alpha=0.03$
and $\eta=10$, $\kappa=2$ in (\ref{flat-profile}). Beam and plasma
transverse temperatures are 1keV.} \label{plot-check-order-1}
\end{figure}

In order to assess the validity of the first-order analytical
results presented above and get an insight into the longer time
evolution of the beam-plasma system, we performed particle-in-cell
(PIC) simulations using the code CALDER \cite{calder} in 2-1/2
dimensions $(x,y,v_{x},v_{y},v_{z})$. All species, namely plasma
ions and electrons and beam electrons, are described as particles.
Beam electrons are injected at $t=0$ in a plasma without current
compensation and $\lambda_{s}=0.05$ $\mu$m. Fig.
\ref{plot-check-order-1} presents the early time evolution of the
beam radial and poloidal mean velocities. The initial beam density
profile was given by Eq. (\ref{flat-profile}) with $\eta=10$ and
$\kappa=2$. The beam was monokinetic with $\gamma_{0}=15$ and
$\alpha$ was equal to 0.03. For beam radial velocity, this figure
shows a nice agreement with the analytical result (\ref{solu_vb1r}).
As for beam poloidal velocity, it is initially vanishing and its
$m=0$ component does remain so. However, poloidal symmetry
eventually breaks due to the arising of filamentation instability.
This takes place after a short transient during which plasma
backcurrent grows. Then the average poloidal velocity start to grow
exponentially, with a growth rate that nicely fits the linear
filamentation instability one, given by
$\gamma_{f}=\beta_{0}\sqrt{\alpha/\gamma_{0}}\omega_{pe}$
\cite{Fainberg70,BFD05}.

\begin{figure}[tbph]
\begin{center}
\resizebox{8.0cm}{!}{\includegraphics{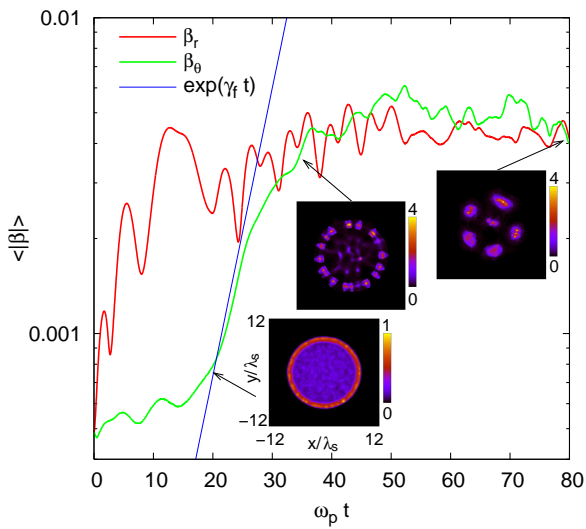}}
\end{center}
\caption{Same as Fig. \ref{plot-check-order-1} for $\alpha=0.15$,
$\gamma_{0}=3$, $\eta=10$ and $\kappa=20$. Transverse snapshots of
beam density are included. Beam and plasma transverse temperatures
are 5 keV.} \label{plot-time-evolution}
\end{figure}

Fig. \ref{plot-time-evolution} presents the longer term evolution of
the transverse components of beam velocity for a monokinetic beam
with a larger value of $\alpha$ ($\alpha=0.15$) and smaller
$\gamma_{0}$ ($\gamma_{0}=3$). As previously, there is an initial
phase, between t=0 and $20 \omega_{pe}^{-1}$, during which radial
velocity grows fast and poloidal velocity remains small. For
$\omega_{pe}t\simeq20$, we can see on the inset of Fig.
\ref{plot-time-evolution} that beam density presents a clear ring
structure at its edge. When the beam current is partially
neutralized, filamentation instability starts, breaking the initial
azimuthal system symmetry and producing the exponential growth of
poloidal beam velocity. When the instability saturates
($\omega_{pe}t\simeq35$), the magnitudes of both components of the
transverse velocity are similar: transverse isotropy is reached. For
$\eta$ larger (not shown here), the relative thickness of the
initial ring diminishes and once filamentation saturates, the
initial structure becomes almost undetectable.

\begin{figure}[tbph]
\begin{center}
\resizebox{8.0cm}{!}{\includegraphics{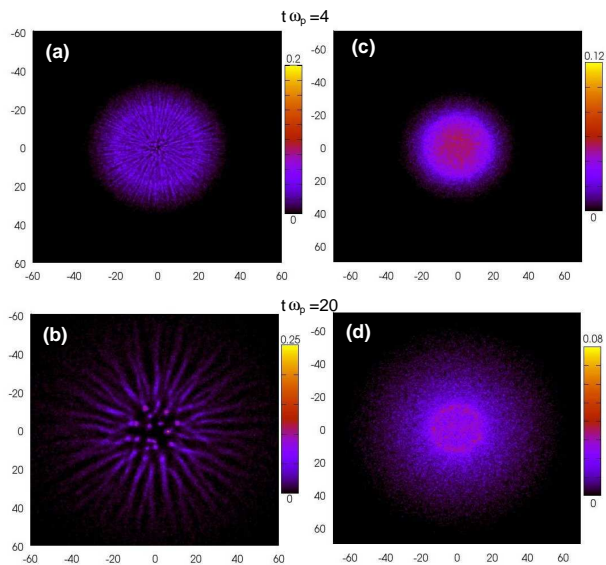}}
\end{center}
\caption{Spatial distribution of the beam density at two times for
beams with angular divergence of $15^{\circ}$. Zero emittance (a-b)
and large emittance (0.25 $\mu$m) cases (c-d) are shown.}
\label{plot-with-divergence}
\end{figure}

Finally, the evolution changes significantly for beams with finite
initial angular divergence. Experimentally, it was found
\cite{Santos02} that electron beams created by focalizing a laser
pulse over a solid target may present divergencies as large as
$17^{\circ}$. The origin of these large divergencies is not clear.
We performed simulations with beams having an angular divergency of
$15^{\circ}$ and two values of emittance. For zero emittance
(laminar beam) a dim ring structure appears at short times (Fig.
\ref{plot-with-divergence}(a)), then filamentation takes place (Fig.
\ref{plot-with-divergence}(b)). The large radial velocity produces a
fast coalescence of the filaments along the radial direction,
resulting in a star-like density pattern. In the high emittance case
(Fig. \ref{plot-with-divergence}(c-d)), no ring structure is
apparent. Moreover, the large transverse temperature prevents the
onset of the filamentation instability \cite{BFD05,Silva02}.

In conclusion, our analysis has shown that, depending on its initial
radial density and velocity distribution, the shape of an electron
beam propagating in a plasma may evolve into a transient ring
structure. This results from the natural evolution of the system and
not from the usually invoked Weibel instability. If its transverse
temperature is low enough, filamentation instability eventually
proceeds. The observation of the ring structure is favored by sharp
beam edges and not too large beam radius (compared to the electron
skin depth). It is not generic which may explain the variety of
experimental observations
\cite{Tatarakis03,Koch2002,Jung2005,SteinReport2003,Norreys}.

Discussions with A. Bret are gratefully acknowledged.


\begin{thebibliography}{99}
\bibitem{note1} It would be more rigourous to introduce the rms radius.
\end{thebibliography}
\end{document}